# Generation of energetic highly elliptical extreme ultraviolet radiation

E. Vassakis[1,2], I. Orfanos[1], I. Liontos[1], D. Charalambidis[1,2,3], P. Tzallas[1,3] and E. Skantzakis[1§]

[1]*Foundation for Research and Technology - Hellas, Institute of Electronic Structure & Laser, PO Box 1527, GR71110 Heraklion (Crete), Greece*
[2]*Department of Physics, University of Crete, PO Box 2208, GR71003 Heraklion (Crete), Greece*
[3]*ELI-ALPS, ELI-Hu Non-Profit Ltd., H-6720 Szeged, Hungary*

§*Corresponding author e-mail address: skanman@iesl.forth.gr*

**Abstract**

In this study the generation of energetic coherent extreme ultraviolet (XUV) radiation with controlled polarization, is reported. The XUV radiation results from the process of high harmonic generation (HHG), in a gas phase atomic medium, driven by an intense two-color circularly polarized counter-rotating laser field, under loose focusing geometry conditions. The energy of the XUV radiation emitted per laser pulse is found to be of the order of ~100 nJ with the spectrum spanning from 17 to 26 eV. The demonstrated energy values along with tight focusing geometries is sufficient to induce nonlinear processes and challenges the perspectives for ultrafast investigations of chiral phenomena in the XUV spectral region.

**Introduction**

Coherent XUV laser-driven sources based on HHG process are extensively explored and applied for a variety of applications in ultrafast science [1-17]. Furthermore, many HHG-based experimental strategies have been implemented in the production and characterization of intense attosecond light sources [15,18-28]. However, these pulsed sources that are as bright as to induce non-linear processes and temporally short enough as to map electronic motion, are mostly limited to emit linearly polarized radiation, being driven by linearly polarized fundamental laser fields. Circularly polarized (CP) XUV light pulses have proven to be a valuable tool and have attracted a keen interest in the scientific community over a broad range of investigations, such as chirality-sensitive light-matter interactions, angle-resolved photoemission spectroscopy and magnetic circular dichroism spectroscopy. Novel approaches in producing CP XUV radiation pulses can be found in the literature [29, 30] but all campaigns so far led to moderate pulse energies.

Nowadays, table-top laser-driven HHG-based sources offer strategies to generate and manipulate the polarization state of highly elliptical/CP XUV light. A key idea underlying several of these methods is to break the symmetry of the system i.e., to either break the symmetry of the emitting medium or the symmetry of the driving field [31-33]. Experimental approaches proposed for the generation of CP XUV radiation involve the exploitation of crossed driving laser beams [34], resonant HHG in elliptical laser fields [35], orthogonally-polarized two-color laser fields [36], bichromatic counter-rotating elliptically polarized drivers [37], circularly polarized counter-rotating fields [38-40] or co-rotating bichromatic laser fields [41,42]. In particular, by implementing counter rotating ω/2ω laser fields, an innovative experimental work back in 1995 reported the generation of polarization dependent high-order harmonics [43]. In this case the polarization state of higher order harmonics can be fully controlled without significantly reducing the conversion efficiency of the HHG process [32,33]. The generated HHG spectrum then results in the generation of a pair of harmonics with opposite helicity. The spectral location of the harmonic orders depends on the wavelength of the fundamental and the second harmonic component. The synthesized driving laser field exhibits a threefold spatiotemporal symmetry. Hence, the 3n-order (where n is a positive integer) harmonics are forbidden in isotropic media, like in the case of an atomic gas phase medium [38,39,44].

The helicity of the 3n+1 and 3n-1 harmonics are the same to those of the fundamental and second harmonic, respectively, and are opposite to each other because of the conservation of spin angular momentum. In the time domain the electric field resulted by the superposition of the harmonics, synthesize an attosecond pulse train in which each pulse is linearly polarized and the polarization axis is rotating by 120$^o$ from pulse to pulse [32,40,45].

Albeit the rich existing literature on the generation of highly elliptical or CP XUV radiation, the energy content of the experimentally generated pulses is so far limited to pJ regime [33,46], mostly due to the focusing geometry used for the HHG. In this work, we apply a bichromatic counter rotating circularly polarized laser field to drive HHG in Argon (Ar) atoms, under loose focusing conditions and demonstrate the production of ~100 nJ highly elliptical polarized XUV radiation in the spectral range of ≈ 20 eV.

**Experimental section**

In producing a bi-circular field for circular polarized HHG a compact MAch-ZEhnder-Less for Threefold Optical Virginia spiderwort-like (MAZEL-TOV-like [47]) scheme is used. A schematic illustration of it is sketched in Figure 1. The experimental investigations were carried out exploiting the MW beamline of the Attosecond Science and Technology Laboratory (AST) at FORTH-IESL. The experiment utilizes a 10 Hz repetition rate Ti:Sapphire laser system delivering pulses up to 400 mJ/pulse energy at $\tau_L$=25 fs duration and a carrier wavelength at 800 nm. The experimental set-up consists of three chambers: the focusing and MAZEL-TOV-like [47] device chamber, the harmonic generation chamber and the detection chamber. All chambers are under vacuum. A laser beam of 3 cm outer diameter and energy of ≈ 20 mJ/pulse is passing through a 3 m focal-length lens with the MAZEL-TOV-like device positioned 1.25 meters downstream. The apparatus consists of a beta-phase barium borate crystal (BBO), a calcite plate and a super achromatic quarter waveplate (Fig. 1). A fraction of the energy of the linear p-polarized fundamental pulse, is converted into a perpendicular (s-polarized) second harmonic field (410 nm) in a BBO (0.2 mm, cutting angle 29.2$^0$ for type I phase matching). The conversion efficiency of the BBO crystal was maximized and it was found ≈ 30%. The run-out introduced by the BBO crystal for the SHG of 800 nm was determined to be 38.6 fs. It is noted that by placing the BBO after the focusing lens ensures that the wavefronts of the converging fundamental laser beam are reproduced into that of the second harmonic field. Therefore, the foci (placed close to a pulsed gas jet filled with Ar) of the ω and 2ω fields coincide along the propagation axis. Additionally, the beam passes through a calcite plate at almost normal incidence (AR coated, group velocity delay (GVD) compensation range 310-450 fs), which pre-compensates group delays introduced by the BBO crystal and the super achromatic quarter waveplate. The super achromatic waveplate converts the two-color linearly polarized pump into a bi-circular field, consisting of the fundamental field and its second harmonic, accumulating at the same time a group delay difference of 253 fs between the 410 nm and 800 nm wavelengths. Assuming Gaussian optics, the intensity at the focus for the two components of the bicircular polarized field is estimated to be $I_\omega \approx I_{2\omega} \approx 1 \times 10^{14}$ W/cm$^2$. After the jet, the produced XUV co-propagates with the bicircular driving fields towards a Si plate, which is placed at 75º reducing the p-polarization component of the fundamental and the second harmonic radiation while reflecting the harmonics [48] towards the detection area. Directly after the Si plate, a pair of 5 mm diameter apertures were placed in order to block the outer part of the ω and 2ω beams, while letting essentially the entire XUV through. A 150 nm thick Sn filter is attached to the second aperture, not only for the spectral selection of the XUV radiation, but also to eliminate the residual bicircular field. A calibrated XUV photodiode (XUV PD) was introduced into the beam path in order to measure the XUV pulse energy. The transmitted beam enters the detection chamber, where the spectral characterization of the XUV radiation takes place. The characterization is achieved by recording the products of the interaction between the XUV generated beam and the gas phase Ar atoms introduced by a pulsed gas jet valve. The electrons produced by the interaction of Ar atoms with the unfocused XUV radiation were detected by a μ-metal shielded time-of-flight (TOF) spectrometer. The spectral intensity distribution of the XUV radiation is obtained by measuring the single-photon ionization photo-electron (PE) spectra induced by the XUV radiation with photon energy higher than the $I_p$ of Ar ($I_{pAr}$=15.76 eV).

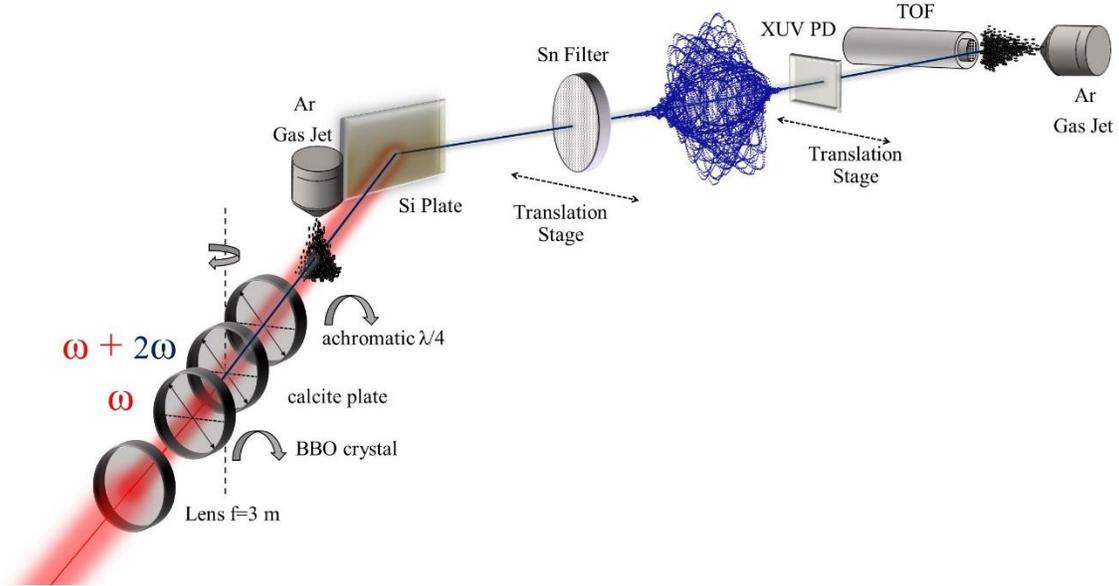

**Figure 1.** Experimental apparatus for the generation of energetic XUV radiation with controllable polarization. A compact (15 cm long) MAZEL-TOV-like device is installed after a 3-m lens. The device includes a BBO crystal, a calcite plate, and a rotatable super achromatic quarter waveplate. The two-color bi-circular field beam is focused into a pulsed gas jet filled with Ar. The generated XUV radiation is reflected towards the detection area by a Si plate. The detection chamber consists of a calibrated XUV photodiode (XUV PD), a pulsed gas jet filled with Ar and finally a μ-metal shielded TOF spectrometer.

Exploiting a MAZEL-TOV-like device, the generated HHG spectrum changes from a spectrum which contains only the odd harmonics, to a spectrum containing 3n±1, n=1,2,3… harmonic orders. This is a typical spectral signature of highly elliptical polarized HHG radiation **[29,32,33]**. Figures 2 (a) and inset, show the recorded HHG spectrum of the highly elliptical XUV radiation generated in Ar gas when no metal filter is used. Additionally, Figure 2 (b) shows the spectra of the XUV radiation transmitted through a 150 nm thick Sn filter. At this point it should be perceived that single photon ionization is a linear process and ionization by highly elliptical polarization can be understood as a simple sum of ionization from two perpendicular linear components of the polarized laser fields. Therefore, the ionization cross section for Argon is assumed to be the same as in the case of linear polarized light **[49]**. The red (blue) circular arrow in Figure 2 indicates the rotation direction of the polarization of the fundamental (second harmonic) used for HHG and the transferred helicity to each harmonic.

The highest harmonic order observed was the 17$^{th}$ for Ar exploited as generating medium. According to the cut-off law, for equal intensities of the ω and 2ω laser field components ($I_\omega = I_{2\omega}$), the highest XUV photon energy emitted is given by the expression: $E_{max} = 1.2\, I_p + \frac{1}{\sqrt{2}}\, 3.17\, U_p$, where $U_p = U_{p\omega} + U_{p2\omega}$ [34,50,51]. Therefore, in the present experimental conditions the cutoff photon energy is expected to be around 23$^{rd}$ harmonic. This deviation is attributed to the macroscopic response of the medium and selective phase matching conditions [33,52] which depend on the relative position of the focus of the driving field and the gas jet. Additionally, because of the p character of the ground state of the Ar gas, the harmonics of one helicity are stronger than the harmonics of the opposite one [53]. Finally, the collection efficiency of the TOF spectrometer is lower (compared to the case of linear polarization) when highly elliptical ionizing radiation is applied [54,55], restricting the highest possible observable harmonic orders.

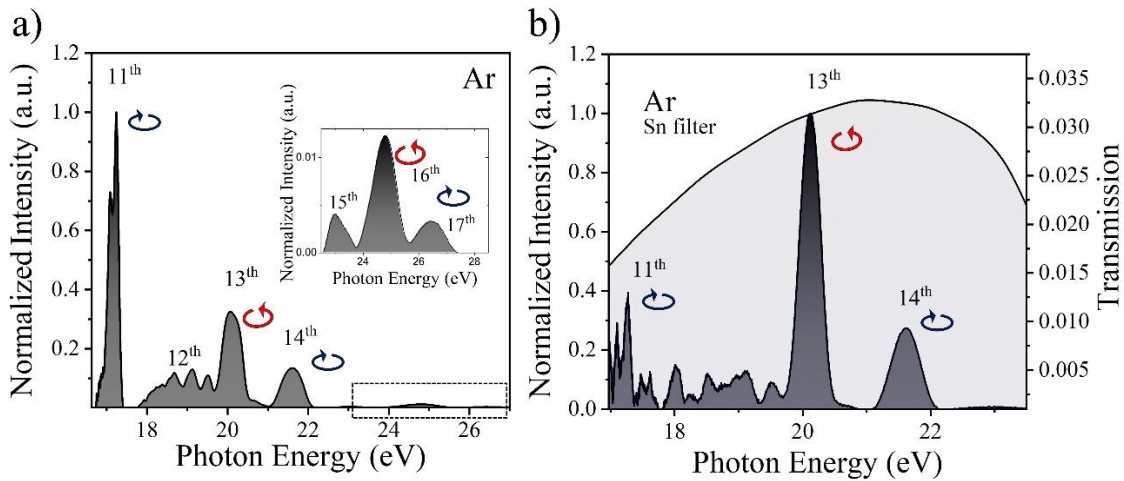

**Figure 2.** (a) Recorded photoelectron spectrum produced by the interaction of unfiltered highly elliptical polarized XUV radiation with Ar gas. Inset shows the magnifying region around harmonic orders 15 to 17 (b) Spectrum as in (a) but produced by the XUV radiation transmitted through a 150 nm thick Sn filter.

At this point it should be stressed that there is an uncertainty considering the degree of circular polarization of the driving laser shaped by the achromatic quarter waveplate, due to the deviation from an ideal waveplate both in retardation and in the crystal axis. Therefore, the symmetry of the superimposed electric field deviated from the threefold symmetry. Thus, the 12$^{th}$ and 15$^{th}$ harmonics are still weakly observed in the spectra. The appearance of the 3n harmonic orders can be induced by the driving fields imperfect overlap [56] or by the anisotropy of the generating medium [57,58], a case which is less probable in this experimental investigation since noble gasses are exploited as the generating media and due to the collinear configuration of the MAZEL-TOV like device. Lou Barreau and colleagues [59] showed that by solving the TDSE breaking can occur even for perfectly circular driving fields and in an isotropic medium, when the strong-field interaction resulting to the harmonic emission presents sub-ω cycle modulations due to (i) a fast temporal variation of the driving laser fields on the envelope rising and falling edges, causing temporal variations of the harmonic dipole vector and (ii) ionization of the medium resulting in a fast decay of the induced dipole strength with time.

Moreover, emphasis should be given to the fact that the harmonic ellipticity and helicity can be fully controlled at the source by simply rotating the fast axis of the super achromatic waveplate [32]. This is based on the conservation of spin angular momentum in HHG process [32]. Considering the formulation described in [32] the equation characterizing the ellipticity for the emitted harmonics allowed by the selection rules as a function of the angle α of the fast axis of the super achromatic waveplate in the case of MAZEL-TOV-like device can be extracted from: $\varepsilon_{n_1,n_2} = \frac{1-\sqrt{1-(n_1-n_2)^2 \sin^2(2\alpha)}}{(n_1-n_2)\sin(2\alpha)}$ where $\varepsilon_{n_1,n_2}$ is the ellipticity of the $(n_1, n_2)$ harmonic channel where $n_2 = n_1 \pm 1$. The equation is valid for ideal overlap of the two foci and the generation by isotropic medium. Figure 3(a) presents the estimated harmonic ellipticity $\varepsilon_{n_1,n_2}$ as a function of the rotation angle α of the super achromatic waveplate in our MAZEL-TOV-like device installed in the MW beamline of AST at FORTH-IESL. Figure 3(b) shows the normalized experimental PE spectra as a function of the angle α of the fast axis of the waveplate in the case of Ar as generating medium.

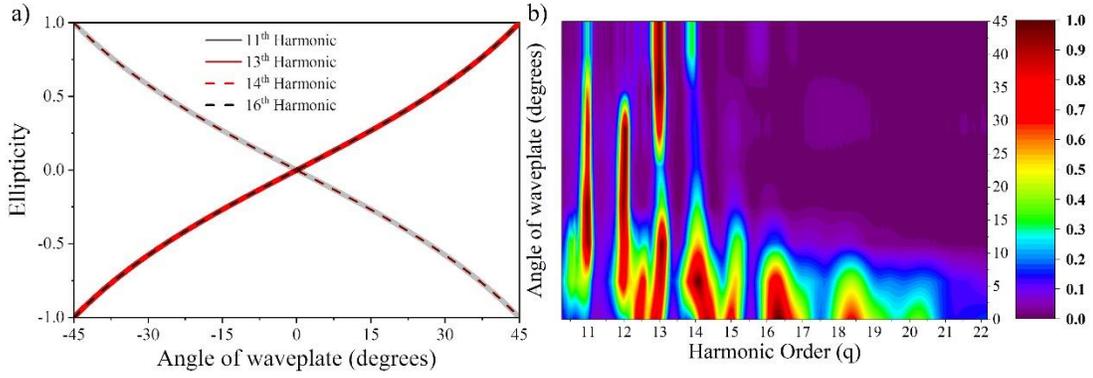

**Figure 3.** (a) Theoretically predicted ellipticity ε of harmonics 11$^{th}$, 13$^{th}$, 14$^{th}$, 16$^{th}$ generated in an isotropic medium as a function of the angle α of the super achromatic waveplate's fast axis (see text). (b) Normalized PE spectra as a function of the angle α of the fast axis of the waveplate in the case of Ar.

**Energy content estimation**

The estimation of the energy content of the highly elliptical XUV radiation emitted per pulse is discussed in detail in this section. The estimation is enabled at a first step by measuring the linearly polarized XUV pulse energy by means of the XUV photodiode. Then, by comparing the HHG spectra depicted in the measured photoelectron spectrum of Ar atoms upon interaction with highly elliptically and linearly polarized XUV light, respectively, we can deduce the energy per pulse of the highly elliptical XUV emission.

The linearly polarized HHG signal recorded with the calibrated XUV photodiode (Opto Diode AXUV100G) placed on the XUV beam path after the Sn filter is fed to an oscilloscope with 50 Ω input impedance and the measured trace was integrated. In Figure 4 (a) a typical measurement of the p-polarized XUV radiation energy is presented. The photodiode signal was measured with the harmonic generation gas jet ON and OFF. The measured signal when the jet was OFF comes from the residual IR radiation and is subtracted from the signal measured when the generation gas was ON. The pulse energy is given by $E_{PD} = \sum_q \frac{n_e \cdot w \cdot h\nu_q}{\eta_q} \cdot e$ where $q$ is the harmonic order, $n_e$ is the number of produced photoelectrons, $w$ is the statistical weight of the qth harmonic, $h\nu_q$ is the harmonic photon energy, $\eta_q$ is the photodiode quantum efficiency [60] and $e$ is the electron charge.

The photoelectron number is given by $n_e = \frac{S_T - S_\omega}{e \cdot R}$ where $S_T$ is the total time integrated photodiode signal, $S_\omega$ is the time integrated photodiode signal when the harmonic generation is OFF, $e$ is the electron charge and R is the oscilloscope impedance. The quantum efficiency of the photodiode as a function of the photon energy is provided by the manufacturing company and it is presented in Figure 4(b). For determining the energy of the harmonic radiation produced at the source one has to consider also the filter transmission as well as the Si reflectivity. Then the produced XUV energy $E$ emitted per laser pulse at the harmonic generation source is given by $E = \sum_q \frac{n_e \cdot w \cdot h\, v_q}{\eta_q \cdot R_q^{Si} \cdot T_q^{Sn}} \cdot e$. Here $T_q^{Sn}$ is the ~3% transmission of the Sn filter in this spectral region measured by recording the harmonic spectrum of linear p-polarized harmonics (with the MAZEL-TOV-like device out of the beam path) with and without filter. The main reason that p-polarized XUV radiation was chosen for the callibration of the Sn filter is the higher signal it results with and without the filter thus minimizing the error of the measurement. $R_q^{Si} = 50\% - 60\%$ is the reflectivity of the Si plate [48]. Under optimal generation conditions the maximum energies at the source were found to be in the range $E_{Ar}^{linear} \approx$ 1µJ. These values concern the emitted 11[th],13[th],15[th],17[th] and 19[th] harmonics laying in the plateau spectral region. Therefore the energy content per harmonic pulse is ~200 nJ at the source for Ar.

A similar measurement for the highly elliptically polarized radiation has a high degree of uncertainty because the difference of the two signals (gas jet ON – gas jet OFF) is rather small. This is due to the fact that when the MAZEL-TOV-like device is introduced in the beam path, the significantly increased amount of light of the fundamental as well as of the second harmonic frequencies reflected from the Si plate, in the case of non-p-polarized fields, prevents an accurate measurement of the highly elliptical XUV using the XUV PD.

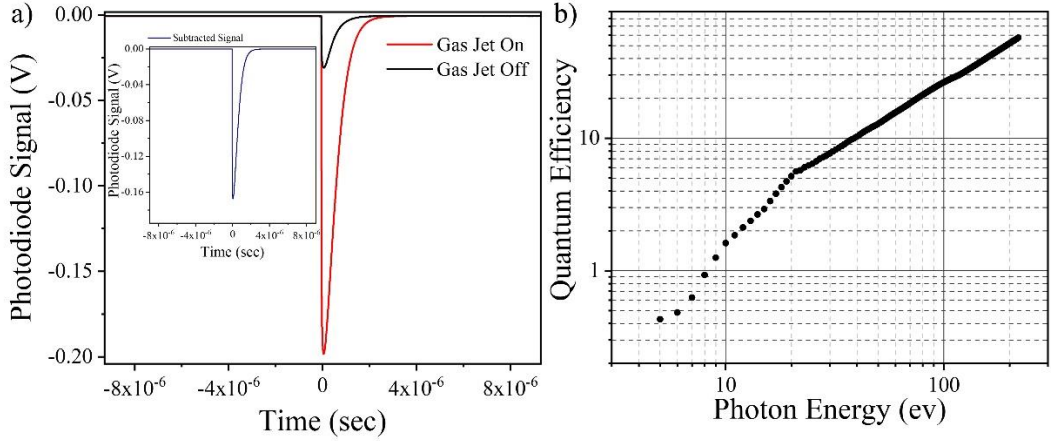

**Figure 4.** (a) Typical measurement of the p-polarized XUV radiation energy in the case of Ar gas. XUV photodiode signal obtained when HHG switched ON (red line) and with the HHG switched OFF (black line). (b) For the extraction of the pulse energy the XUV photodiode quantum efficiency as a function of photon energy provided by the manufacturing company Opto Diode Corp was used.

In order to deduce an estimation of the energy per pulse of the highly elliptical XUV light, in the present configuration, it is necessary to perform measurements of the Ar single photon ionization PE spectra induced by linear p-polarized and highly elliptical XUV radiation, at the same detection conditions, and directly compare them after optimization of harmonic emission in both cases. In Figures 5 and inset the PE spectra in the case of linear and highly elliptical polarization is shown for Ar gas.

Having determined the energy values in the case of linear polarized XUV field and having deduced also the ratios of the harmonic integral values in the case of p-polarization and highly elliptical XUV radiation, by the recordings of the PE spectra, one can extarct the energy content of the latter. The ratio of the integral values $I_q^{highlyelliptical}/I_q^{linear}$ ,(where q is the harmonic order) was found 0.15, 0.17, 0.02, 0.006 for the 11$^{th}$,13$^{th}$ ,14$^{th}$ and 16$^{th}$ harmonic, respectively. This ratio of the integral values of the two different PE spectra (highly elliptical vs linearly polarized) equals to the ratio of the energy content of the, per laser pulse, generated XUV radiation $I_q^{highlyelliptical}/I_q^{linear} = E_q^{highlyelliptical}/E_q^{linear}$ in both polarizations and has to be divided by the correction parameter ~0.6 introduced by the different percentage of the total photoelectons' number entering the TOF spectrometer. This consideration has to be taken into acount due to the different angular photoelectron destributions resulting from the single photon ionization by the two polarization's states (linear, circular) [**54,55**]. After this correction, the energy per laser pulse emitted in the case of highly elliptical radiation was then estimated to be $E_{Ar}^{highly\ elliptical} \approx 100$ nJ. It has to be pointed out that most of the energy is almost equally distributed between $11^{th}$ and $13^{th}$ harmonics and only a small percentage in $14^{th}$ and $16^{th}$ harmonics. Conclusively, the enegy of highly elliptically polarized XUV radiation is ~ 10 times less as compared to linearly polarized XUV radiation at the same spectral region under conditions where the yield was optimized in both cases.

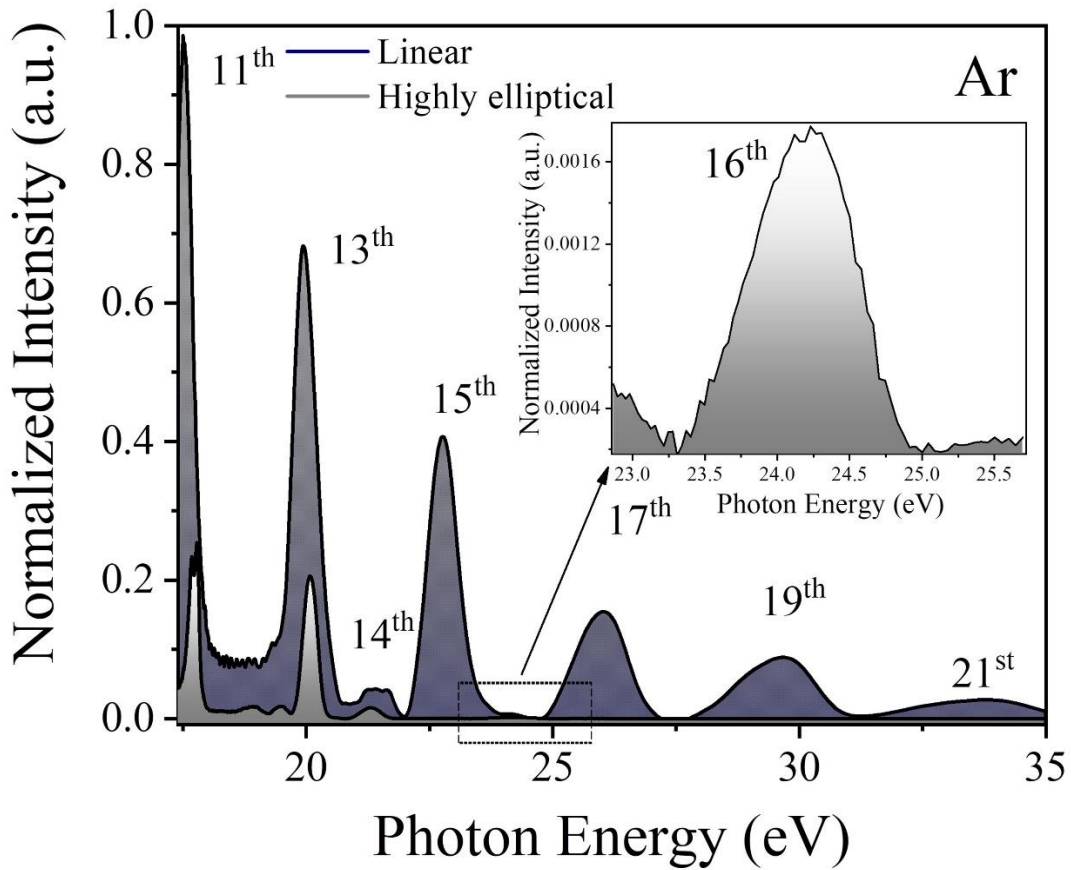

**Figure 5.** PE spectra of Ar induced by the linear p-polarized and highly elliptical XUV radiation generated by Ar at the same detection conditions, as soon as the optimization of harmonic emission is realized in both cases.

The estimated energy content of $E_{Ar}^{highly\ elliptical} \approx$ 100 nJ at the source, per driving laser pulse, in conjunction with the implementation of broadband optical elements and tight focusing configurations is sufficient to induce nonlinear phenomena at the target's area. More specifically, the Si plate used in these experimental investigations can be substituted with custom built multilayers mirrors offering reflectivity up to ~75% in a broad spectral region. A 100 nm thick Al filter can provide transmission of ~80% to the spectral region > 17 eV. Finally, a set of gold coated toroidal mirrors, at Wolter configuration [61] to minimize the coma aberration, can provide high reflectivity (~80%) and a focal spot ~3 μm. The combination of the above-mentioned optical elements and arrangements can result to intensities up to ~$10^{13}$ W/cm$^2$ at the focus.

Reducing the ellipticity of the generated radiation (by varying the fast axis of the super-achromatic waveplate) higher pulse energies were measured. This is a counterplay between the generated energy and the ellipticity of the harmonics.

**Conclusions**

By exploiting the implemented linearly polarized megawatt XUV beamline at FORTH-IESL, we report a method to produce energetic highly elliptical XUV light. The approach is based on gas-phase HHG driven by an intense two-color circularly polarized counter-rotating driving laser field, produced by a MAZEL-TOV-like device when introduced in the linear MW XUV beamline, under loose focusing conditions. The energy content for the highly elliptical XUV light is deduced by direct comparison between the linear and highly elliptical XUV spectra measured in the photoelectron spectra of Ar atoms, once the energy content of the linearly polarized XUV light is determined. The energy per driving laser pulse in the spectral region between 17 to 26 eV, is found to be in the range of ~100 nJ. Tight focusing of this light in conjunction with appropriately broadband optical elements is anticipated to lead to ~$10^{13}$ W/cm$^2$ intensities in the focal area. Such intensities would be sufficient to induce nonlinear phenomena in chiral systems in the XUV spectral region.

**Acknowledgments**

We acknowledge support of this work by "HELLAS-CH" (MIS Grant No. 5002735) [which is implemented under the "Action for Strengthening Research and Innovation Infrastructures," funded by the Operational Program "Competitiveness, Entrepreneurship and Innovation" (NSRF 2014–2020) and co-financed by Greece and the European Union (European Regional Development Fund)], the European Union's Horizon 2020 research ELI-ALPS is supported by the European Union and cofinanced by the European Regional Development Fund (GINOP Grant No. 2.3.6-15-2015-00001), the LASERLAB- EUROPE (EU's Horizon 2020 Grant No. 871124), the IMPULSE project Grant No. 871161), the Hellenic Foundation for Research and Innovation (HFRI) and the General Secretariat for Research and Technology (GSRT) under grant agreements [GAICPEU (Grant No 645)] and NEA-APS  HFRI-FM17-3173.

**References**

[1] Haight R and Seidler P F 1994 *Appl. Phys. Lett.* 65 517–9

[2] Haessler, S., Caillat, J., Boutu, W. et al. 2010 *Nature Phys.* 6 200–206

[3] Baker S, Robinson J S, Haworth C A, Teng H, Smith R A, Chirilă C C, Lein M, Tisch J W G and Marangos J P 2006 *Science* 312 424–7

[4] Cavalieri A L et al 2007 *Nature* 449 1029–32

[5] Sandberg R L et al 2007 *Phys. Rev. Lett.* 99 098103

[6] Miaja-Avila L, Saathoff G, Mathias S, Yin J, La-o-vorakiat C, Bauer M, Aeschlimann M, Murnane M M and Kapteyn H C 2008 *Phys. Rev. Lett.* 101 046101

[7] Li W, Zhou X, Lock R, Patchkovskii S, Stolow A, Kapteyn H C and Murnane M M 2008 *Science* 322 1207–11

[8] Goulielmakis E et al 2010 *Nature* 466 739–43

[9] Seaberg M D et al 2011 *Opt. Express* 19 22470–9

[10] Mathias S et al 2012 *Proc. Natl Acad. Sci. USA* 109 4792–7

[11] Hoogeboom-Pot K M et al 2015 *Proc. Natl Acad. Sci. USA* 112 4846–51

[12] Miao J, Ishikawa T, Robinson I K and Murnane M M 2015 *Science* 348 530–5

[13] Zhang B, Gardner D F, Seaberg M D, Shanblatt E R, Kapteyn H C, Murnane M M and Adams D E 2015 *Ultramicroscopy* 158 98–104

[14] E. Skantzakis, P. Tzallas, J. E. Kruse, C. Kalpouzos, O. Faucher, G. D. Tsakiris, and D. Charalambidis 2010 *Phys. Rev. Lett.* 105 043902

[15] Tzallas P, Skantzakis E, Nikolopoulos L et al. 2011 *Nature Phys.* 7 781–784

[16] S. Chatziathanasiou, I. Liontos, E. Skantzakis, S. Kahaly, M. Upadhyay Kahaly, N. Tsatrafyllis, O. Faucher, B. Witzel, N. Papadakis, D. Charalambidis and P. Tzallas 2019 *Phys. Rev.* A 100 061404(R)

[17] R. Eramo, S. Cavalieri, C. Corsi, I. Liontos and M. Bellini 2011 *Phys. Rev. Lett.* 106 213003

[18] Takahashi E., Lan P., Mücke O. et al 2013 *Nat Commun.* 4 2691

[19] Fu Y., Nishimura K., Shao R. et al 2020 *Commun. Phys.* 3 92

[20] Nayak A., Orfanos I., Makos I., Dumergue M., Kühn S., Skantzakis E., Bodi B., Varju K., Kalpouzos C., Banks H. I. B., Emmanouilidou A., Charalambidis D. and Tzallas P. 2018 *Physical Review* A 98(2) 66

[21] Makos I., Orfanos I., Nayak A., Peschel J., Major B., Liontos I., Skantzakis E., Papadakis N., Kalpouzos C., Dumergue M., Kühn S., Varju K., Johnsson P., L'huillier A., Tzallas P. and Charalambidis D. 2020 *Scientific reports* 10(1) 3759

[22] B Senfftleben, M Kretschmar, A Hoffmann, M Sauppe, J Tümmler, I Will, T Nagy, M J J Vrakking, D Rupp and B Schütte 2020 *J. Phys. Photonics* 2 034001

[23] J.-F. Hergott, M. Kovacev, H. Merdji, C. Hubert, Y. Mairesse, E. Jean, P. Breger, P. Agostini, B. Carré, and P. Salières 2002 *Phys. Rev.* A 66 021801(R)

[24] Manschwetus B., Rading L., Campi F., Maclot S., Coudert-Alteirac H., Lahl J., Wikmark H., Rudawski P., Heyl C. M., Farkas B., Mohamed T., L'huillier A. and Johnsson P. 2016 *Physical Review* A 93(6)

[25] S. Chatziathanasiou et al 2017 *Photonics* 4 26

[26] P. Tzallas et al 2005 *J. Mod. Opt.* 52 321-338

[27] Y. Nomura et al 2009 *Nat. Phys.* 5 124

[28] R. Horlein et al 2010 *New J. Phys.* 12 043020

[29] J Schmidt, A Guggenmos, M Hofstetter, S H Chew and U Kleineberg 2015 *Opt. Express* 23 33564-33578

[30] B. Vodungbo, A. Barszczak Sardinha, J. Gautier, G. Lambert, C. Valentin, M. Lozano, G. Iaquaniello, F. Delmotte, S. Sebban, J. Lüning, and P. Zeitoun 2011 *Opt. Express* 19(5) 4346–4356

[31] Skantzakis E., Chatziathanasiou S., Carpeggiani P. et al 2016 *Sci. Rep.* 6 39295

[32] Fleischer A., Kfir O., Diskin T. et al 2014 *Nature Photon.* 8 543–549

[33] Kfir O., Grychtol P., Turgut E. et al 2015 *Nature Photon.* 9 99–105

[34] Hickstein D D et al 2015 *Nat. Photon.* 9 743–50

[35] Ferré A et al 2015 *Nat. Photon.* 9 93–8

[36] Lambert G et al 2015 *Nat. Commun.* 6 6167

[37] Fleischer A, Sidorenko P and Cohen O 2013 *OSA Tech. Dig.* QW1A.6

[38] Long S, Becker W and McIver J K 1995 *Phys. Rev.* A 52 2262–78

[39] Milošević D B, Becker W and Kopold R 2000 *Phys. Rev.* A 61 063403

[40] Milošević D B and Becker W 2000 *Phys. Rev.* A 62 011403

[41] Zuo T and Bandrauk A D 1995 *J. Nonlinear Opt. Phys. Mater.* 04 533–46

[42] Bandrauk A D and Lu H 2003 *Phys. Rev.* A 68 043408

[43] Eichmann H, Egbert A, Nolte S, Momma C, Wellegehausen B, Becker W, Long S and McIver J K 1995 *Phys. Rev.* A 51 R3414–7

[44] Möller M, Cheng Y, Khan S D, Zhao B, Zhao K, Chini M, Paulus G G and Chang Z 2012 *Phys. Rev.* A 86 011401

[45] Fan T et al 2015 *Proc. Natl Acad. Sci. USA* 112 14206–11

[46] A Comby *et al* 2020 *J. Phys. B: At. Mol. Opt. Phys.* 53 234003

[47] O Kfir, E Bordo, G I Haham, O Lahav, A Fleischer and O Cohen 2016 *Appl. Phys. Lett.* 108 211106

[48] E. J. Takahashi, H. Hasegawa, Y. Nabekawa and K. Midorikawa 2004 *Opt. Lett.* 29 507

[49] Majety V.P. and Scrinzi A. 2015 *Photonics* 2 93-103

[50] D. B. Milošević, W. Becker, R. Kopold and W. Sandner 2001 *Laser Physics* Vol. 11 No. 2 pp. 165–168

[51] E. Hasović, W. Becker and D. B. Milošević 2016 *Opt. Express* 24 6413-6424

[52] O Kfir et al 2016 *J. Phys. B: At. Mol. Opt. Phys.* 49 123501

[53] Dejan B. Milošević 2015 *Opt. Lett.* 40 2381-2384

[54] Cui Hui-Fang and Miao Xiang-Yang 2020 *Chinese Physics B* 29(7) 074203

[55] Katharine L. Reid 2003 *Annual Review of Physical Chemistry* 54:1 397-424

[56] Jiménez-Galán A., Zhavoronkov N., Schloz M., Morales F. and Ivanov M. 2017 *Opt. Expr.* 25 22880–22896

[57] Baykusheva D., Ahsan M., Lin N. and Wörner H. 2016 *Phys. Rev. Lett.* 116 123001

[58] Yuan K.-J. and Bandrauk A. D. 2018 *Phys. Rev.* A 97 023408

[59] Barreau L., Veyrinas K., Gruson V. et al 2018 *Nat. Commun.* 9 4727

[60] M. Krumrey and E. Tegeler, R. Goebel and R. Köhler 1995 *Review of Scientific Instruments* 66 4736

[61] H. Wolter 1952 *Ann. Phys.* 445 94